\documentclass[conference]{IEEEtran}

\usepackage{setspace,graphicx,amsmath,float,subfig,cite, wrapfig, algorithm, amssymb, multirow,rotating,eurosym,enumerate,algpseudocode,url}

\begin{document}
\title{Big Data in Critical Infrastructures Security Monitoring:
Challenges and Opportunities}

\author{\IEEEauthorblockN{L. Aniello$^{2}$, A. Bondavalli$^{3}$, A. Ceccarelli$^{3}$, C. Ciccotelli$^{2}$, M. Cinque$^{1}$, F. Frattini$^{1}$,\\ A. Guzzo$^{4}$, A. Pecchia$^{1}$, A. Pugliese$^{4}$, L. Querzoni$^{2}$, S. Russo$^{1}$\\}
\IEEEauthorblockA{\\
(1) DIETI, Univ. of Naples Federico II, Via Claudio 21, Naples, Itlay --  \textit{macinque@unina.it}\\
(2)  Univ. of Rome ``La Sapienza'', Cyber Intelligence and Information Security Res. Center, Via Ariosto 25, Rome, Italy\\
(3) Univ. of Florence, Dept. of Mathematics and Informatics, Viale Morgagni 65, Florence, Italy\\
(4) DIMES, Univ. of Calabria, Via Bucci, Rende, Italy -- \textit{\{guzzo,apugliese\}@dimes.unical.it} 
}
}

\maketitle

\begin{abstract}
%\boldmath
Critical Infrastructures (CIs), such as smart power grids, transport systems, and financial infrastructures, are more and more vulnerable to cyber threats, due to the adoption of commodity computing facilities. Despite the use of several monitoring tools, recent attacks have proven that current defensive mechanisms for CIs are not effective enough against most advanced threats. 
In this paper we explore the idea of a framework leveraging multiple data sources 
%, not always considered by current CI monitors, 
to improve protection capabilities of CIs. 
Challenges and opportunities are discussed 
%and a framework for addressing them is proposed.
along three main research directions: 
i) use of distinct and heterogeneous data sources, 
ii) monitoring with adaptive granularity, and 
iii) attack modeling and runtime combination of multiple data analysis techniques.
\end{abstract}

%\IEEEpeerreviewmaketitle

\section{Introduction}
% ***** START UNINA
Over the past years attackers' community has developed smarter worms and rootkits to achieve a variety of objectives, which range from credentials compromise to sabotage of physical devices. Cyber threats are targeting extremely diverse critical application domains including e-commerce systems, corporate networks, datacenter facilities and industrial systems.  %For example, 
%Stuxnet, i.e., a recent worm discovered in July 2010 and firstly affecting Iranian nuclear plants, has shown that attackers might physically damage critical infrastructures in the near future. 
For example, on July 2010, the well known Stuxnet~\cite{stuxnet_symantec_dossier} cyber attack was launched to damage gas centrifuges located at the Natanz fuel enrichment plant in Iran by modifying their speed very quickly and sharply.
On August 2012, the Saudi oil giant Aramco was subjected to a large cyber attack\footnote{\url{http://www.net-security.org/secworld.php?id=13493}} that affected about 30000 workstations.
On April 2012, the big payment processing provider Global Payments confirmed a massive breach\footnote{\url{http://www.net-security.org/secworld.php?id=12680}} that compromised about 1.5~million cards.
On January 2013, the U.S. Department of Energy underwent an intrusion\footnote{\url{http://www.net-security.org/secworld.php?id=14353}} to 14 of its servers and many workstations located at the Department's headquarters, aimed at exfiltrating personal information about its employees.

%\textbf{Protecting critical infrastructures (CI)} has thus become a compelling need; moreover, the increasing adoption of general purpose computing equipment in CIs, such as HW, operating systems, and applications, makes them inherently vulnerable to known and zero-day exploits, intentional misuse, or bad operator practices (such as plugging in an infected media or writing down a password). 

Analysis of data collected under real workload conditions plays a key role to monitor system activities and to detect ongoing anomalies. 
%While much work exists in the area of reliability data analysis since early computer systems, security data analysis still poses a number of research challenges and opportunities.
CIs are currently equipped with a variety of monitoring
tools, such as \textit{system and application logs}, \textit{intrusion
detection systems (IDS)}, and \textit{network monitors}. However, recent cyber attacks have proven that today's defensive mechanisms are not effective against most advanced threats.
%it is still not clear how these data sources and related analysis techniques are able to cope with more recent cyber threats. 
For example, Stuxnet was able to fool the supervisory control and data acquisition (SCADA) system by altering the reading of sensors deployed on the centrifuge engines, and it went undetected for months. %This worm could have been detected by sampling actual signals sent to sensors and actuators. %Similarly, credential compromise attacks or insider threats can go undetected by regular IDS due to the use of legitimate credential by the attackers: the adoption of historical user profiles is potentially useful to pinpoint ongoing suspicious activity. Examples show that effective security analysis should involve heterogeneous and large volumes of data sources that are not always collected by current CI monitors.

Among the possible countermeasures that could be adopted, the idea of leveraging distinct and heterogeneous data sources can help to draw a clearer picture of the system to protect. Indeed, by correlating diverse information flows coming from multiple origins not always collected by current CI monitors, it can be possible to extract additional insights on potentially threatening activities that are being carried out. For instance, the presence of Stuxnet could possibly have been detected by \emph{monitoring several other operational and environmental parameters}, like the centrifuge energy consumption, and by correlating their readings to infer possible anomalies in the status \cite{6161536} (e.g., fluctuating power consumption in a centrifuge, correlated with a stable rotational speed can be considered as an anomalous state). In addition, according to a CyberArk's report~\cite{cyberark_report_2013}, several successful attacks including the ones reported above exploited privileged accounts to achieve their objectives, and the same report states that ``86\% of large enterprises (across North America and EMEA) either do not know, or have grossly underestimated the magnitude of their privileged account security problem''.
A possible solution could consist in leveraging the \emph{monitoring of the activities of such privileged accounts} to pinpoint ongoing suspicious activity. %with the aim of detecting anomalies that could reveal potentially malicious actions.

The use of multiple and diverse sources producing huge amounts of data calls for the research of new solutions for monitoring and analysis, able to timely and efficiently recognize ongoing malicious activities in CIs.
This paper introduces the basic notions of a framework for data-driven security monitoring and protection of CIs. Our proposal stems from needs and challenges for effective security monitoring and describes an architectural solution to them, moving along the following research directions: i) the use of large amount of data collected from distinct and heterogeneous data sources; ii) the adoption of monitoring strategies with an adaptable level of granularity, %to change at runtime the type and quantity of variables to be monitored depending on the current alert status of the system, hence addressing 
to face the issue of big data volumes; iii) the formalization of attack models and the combination of diverse state-of-art data analysis techniques to improve the capability of detecting threats and triggering protection actions.

%\textbf{TODO} \textit{The rest of the paper is organized as follows}
% ***** END UNINA

\section{Needs and Challenges}

\subsection{Multiple Data Sources}
%We need multiple data sources to have a clear picture of the system under analysis. To detect both IT attacks and cyber-physical attacks stux.net like.

%Description of scenarios where the use of multiple sources can be useful.  What these sources should be (operational, environmental, ... here UNINA can add something in a second round).
%And with what monitoring systems are collected today (analysis of the existing technologies).

%Here the challenge is the size and heterogeneity of (big) data. (some estimates from real installations).

%-----INIZIO CIS----

The idea of using distinct and heterogeneous data sources available in today's CIs can help to draw a clearer picture of the system to protect and of the threatening activities being carried out. The aim is to improve the protection of future CIs exploiting the (hidden) value of data: they are already available but not fully exploited in today CIs.

However, as the size and complexity of systems increase, the amount of information that can be collected by data sources skyrockets. For example, in the 1300-nodes data center we target as case study (see Section \ref{sec:casestudy}) the monitoring system produces about 16.6 GB of data per day, with observed traffic peaks of about 240000 pkt/s.
This is a consequence of multiple factors: \textit{(i)} the increasing availability of cheap HW probes, \textit{(ii)} the ubiquitousness of communication infrastructures (either wired or wireless) and the Internet, and \textit{(iii)} the novel algorithmic approaches that todays make handling huge amounts of data practical. A further important aspect is that the heterogeneity of collected data is going to increase as well: new data sources are connected to monitoring systems to collect and analyze different kinds of data as this could potentially provide useful insights on current system statuses.

This mix of factors marks the shift from a mostly human-controlled distributed monitoring model (think, for example, about how railway companies in the past controlled the status of their infrastructures through hundreds of people deployed on the territory along their tracks to locally monitor and then report to their bosses) to fully automated IT infrastructure for monitoring that tries to relieve as much as possible from humans the burden of analyzing data to infer high-level information. Making this new model practical in scenarios where huge amounts of heterogeneous data are available calls for the research of new algorithmic and architectural solutions able to withstand these new challenges.

%-----FINE CIS----

\subsection{Monitoring with different granularity}

An accurate tuning of the amount of variables to be monitored and the frequency of data collected from system probes appears fundamental to study and plan at design time the computational load on the monitoring infrastructure.

First, it is necessary to select what sources are worth monitoring amongst the many available, considering the target system and also the expected workload. For example in \cite{BovenziBRB11} sources at the OS level, such as amount of free memory, disk throughput, or network throughput, are selected out of hundreds of possible indicators; their relevance for anomaly detection is further explored and confirmed in \cite{bondavalli2013experimental}.

Appropriate selection of data sources is relevant but unfortunately may not be sufficient. In large-scale critical infrastructures, given the number of components, we can reasonably consider that monitoring each parameter using the best possible resolution system is unfeasible. Thus it may be required to define monitoring strategies that minimize the amount of data to analyze and consequently the monitoring resources to be used, still without decreasing the efficacy of the monitor, e.g., adopting different monitoring granularities depending on the current alert level of the system and of its components.
This calls for the definition of new solutions able to find the right compromise in terms of the monitoring grain without having a negative impact on the monitoring accuracy as well as without depleting the resources devoted to monitoring.

%--------- UNIFI ends ------%

\subsection{On-Line Big Data Processing}
%To correlate data and produce alerts on-line by comparing analyzed data with attack models.

%Here the challenges are the formalization of attack models and conformance checks and/or ways to detect deviation from the normal behavior (here UNINA can add something in a second round).

%%%%%%%%%% INIZIO UNICAL %%%%%%%%%
%Further critical challenges arise in those scenarios where 
%a huge amount of data are logged about all the events that happen
%in a critical infrastructure, and it is necessary to look for any part 
%of the log that may represent a (possibly complex) critical event for which
%an alert has to be produced.

The large number of collected data also implies difficulties in the data processing phase.
Several \textit{techniques and tools} have been proposed to analyze raw data with the objective of detecting on-going attacks. However, the performance of the detection, in terms of coverage and false alarm rate, strictly depends on the adopted technique. Solutions which encompass the (on-line) combination of multiple analysis techniques need to be investigated, in order to improve the capability of detecting potential threats and triggering protection actions on the CI.
Recent studies have also proven the usefulness of (temporal and/or typed) \textit{graph-based attack models} \cite{IRI07,ESORICS11,CISIM11,GKR13,TKDE13,TKDE14,cotroneo13}. 
%In these scenarios, several important problems emerge:
%\begin{enumerate}
%\item The \emph{formalization of attack models}. Recent studies have proven
%the usefulness of (temporal and/or typed) graph-based attack models~\cite{IRI07,ESORICS11,CISIM11,GKR13,TKDE13,TKDE14}. 
If we assume that
the input log is a sequence of events having a type and a timestamp,
a graph-based attack model has event types as vertices and is defined in such a way that the paths from start to terminal
vertices in the model represent a critical event/attack when they correspond to subsequences of the log. 
Such subsequences are also called \emph{instances} of the attack model.
%\item \emph{Conformance checking}. 
However, finding correlations among data by comparing analyzed data with attack models 
and producing alerts in an on-line fashion may become extremely difficult when the number of attack models at hand
and the size of the input log increase. It is therefore important 
to ensure the scalability of the algorithms and data structures 
used when performing the \textit{conformance checking} task.
%%%%%%%%%% FINE UNICAL %%%%%%%%%
% ***** START UNINA
%\item The \emph{combination of multiple analysis techniques}. 

%\end{enumerate}
% ***** END UNINA

\section{A framework for Data-Driven Security of CIs}

\begin{figure}[t]
\centering
\includegraphics[scale=0.4]{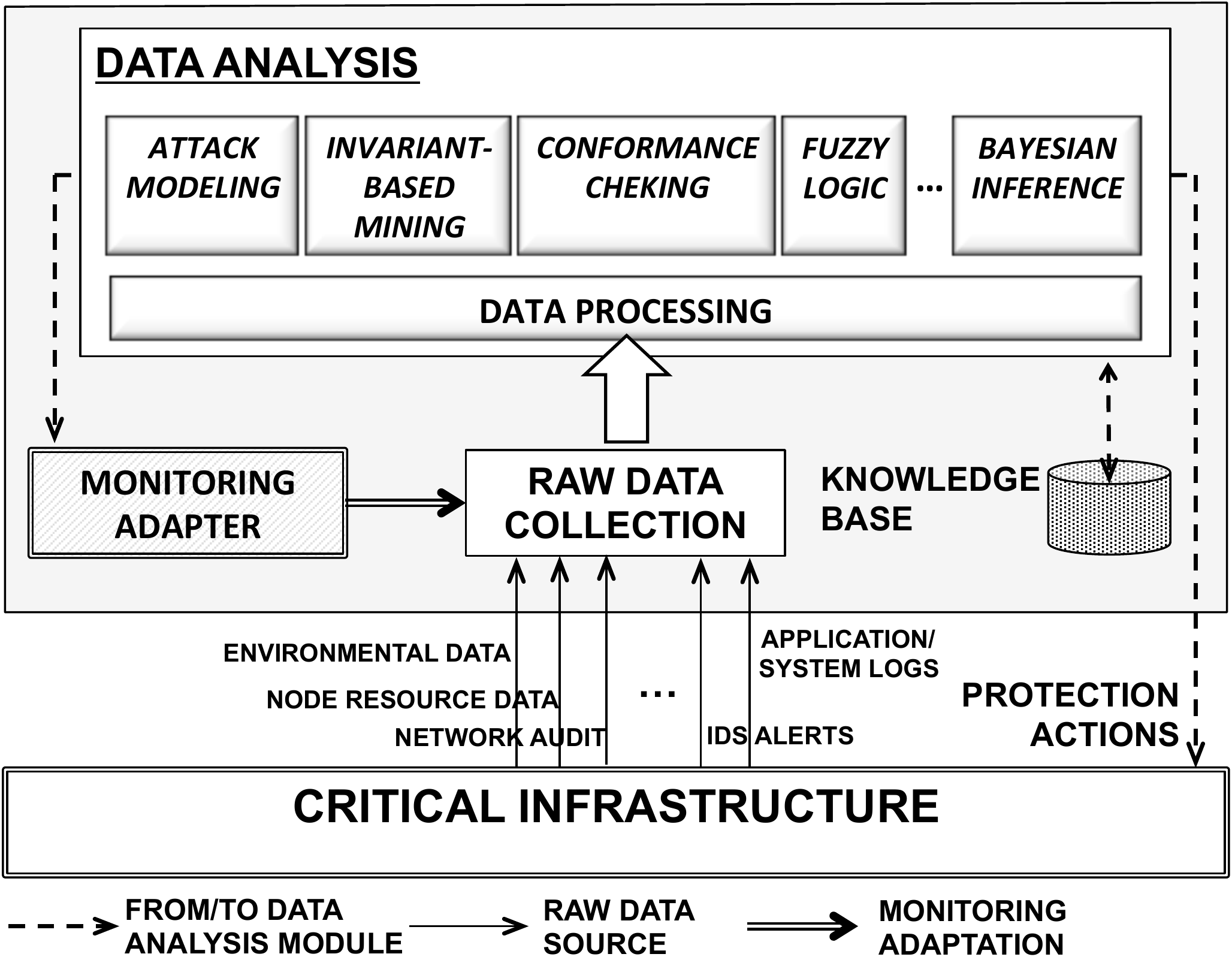}
\caption{\small{The Framework for Data-Driven Security Monitoring and Protection of CIs}}
\label{fig:framework}
\vspace{-0.5cm}
\end{figure}

Figure \ref{fig:framework} proposes an architectural solution to the discussed challenges. The key idea is to combine several data sources and different data analysis techniques to improve the capability of detecting potential threats and triggering protection actions on the CI. The results of the analysis are also useful to assess the current alert level of the CI's components and to adapt the grain of monitoring through the Monitoring Adapter, e.g., to intensify the monitoring of components deemed of suspicious activity and reduce the monitoring of the other ones. The main blocks of the framework are described in the following.

\subsection{Raw Data Collection}
\label{sec:collect}
As the name suggests, the Raw Data Collection block is responsible for gathering raw data from the monitored CIs, exploiting available data monitoring technologies and/or logs produced by diverse software layers or hardware controllers.

%-----INIZIO CIS ----
%Data monitoring involves two different issues: data collection and data analysis.
Many technologies for data monitoring have been developed over the past thirty years, ranging from relatively simple data collection tools (such as Unix syslog\footnote{\url{http://www.ietf.org/rfc/rfc3164.txt}}) to more sophisticated data analysis systems.

Ganglia\footnote{\url{http://ganglia.sourceforge.net/}} is a scalable distributed monitoring system. Its hierarchical design is primarily targeted at federation of clusters and grid computing systems. Each monitored node multicasts its monitored metrics to all nodes in its cluster, enabling automatic discovery of nodes.
%To aggregate the state of multiple clusters Ganglia uses a tree of point-to-point connections. Each leaf of the tree is a representative node of a distinct cluster, while non-leaf nodes aggregate the state of child nodes. Clients can poll each node in the tree to get the desired monitored data.
Nagios\footnote{\url{http://www.nagios.org/}} is one of the most used open source monitoring systems. It offers an advanced notification system and is extensible through plug-ins. Its functioning is based on both active and passive checks of services, hosts and network state. %It relies on a hierarchical description of the system to determine if an host/service is down or unreachable. 
Splunk\footnote{\url{http://www.splunk.com/}} is a commercial monitoring system that allows search, filtering and analysis of structured and unstructured textual logs through indexing.
Artemis~\cite{Cretu-Ciocarlie:2008:HPA:1855886.1855888} is a monitoring system primarily designed for analyzing large-scale distributed logs. It has a modular design, separating data collection from data analysis. The log collection module supports heterogeneous data sources and types (e.g. text, binary, XML). Collection and analysis modules are extensible through plugins and application-specific functions. %before being stored in a relational database. The data analysis module is extensible through plugins and supports generic and user-defined functions.
Chukwa\footnote{\url{https://chukwa.apache.org/}} is a MapReduce-based log collection and monitoring system. It uses the Hadoop distributed file system (HDFS) as a store and analyzes collected logs through MapReduce jobs. It is designed to collect data from hundreds of sources reducing the number of required HDFS files.
%Since HDFS has been designed for batch processing rather than continuous processing, the design of Chukwa is primarily aimed at adapting the typical workload of a monitoring system (with multiple sources producing a huge number of relatively small files and low-rate concurrent writes) to the HDFS characteristics (that make HDFS particularly suited for working with large files at high write rates).
%The architecture of Chuckwa is based on collectors processes that receive data from hundreds of sources. Each collector writes all data to a single sink file. This approach drastically reduces the number of generated HDFS files, making the workload more suitable for HDFS. 
%Moreover, since HDFS writes are not visible until the file is closed, collectors periodically close their file and create new ones. Closed files are thus available for processing by a MapReduce job. This clearly limit the analysis of monitored data to a timescales of minutes.
%-----FINE CIS ----
Logbus\footnote{\url{http://www.critiware.com/logbus.html}} is a framework for the collection and analysis of rule-based logs, i.e., logs produced according to formal rules in the source code and designed to improve the detection of runtime problems in terms of detection rate and false alarms~\cite{Logbus}.

% ***** START UNINA

%Monitoring systems for large data centers can collect data from a number of sensors and about several aspects of the system.
Data collectable through these monitoring systems can be classified in three broad categories: performance, environment, and operational data.
\textbf{Performance data} are among the most monitored, and are related to the use of system resources. 
Main examples are the usage of CPU or memory. Other sources are about the use of the network, such as inbound and outbound traffic.
\textbf{Environment data} are rarely exploited, even if they could be very useful to detect ongoing cyber physical attacks. They describe the status of the environment in which the system is placed and nodes' parameters not related to performed activities.
In this category fall temperature, humidity, the status of cooling systems, if any, etc. 
The monitoring of the energy consumption is also in this category.
%It can be measured for each node in the system or for group of nodes (e.g., energy consumption of a rack).
Finally, \textbf{Operational data} encompass all the information achieved by collecting, and presumably parsing and filtering, logs from the various software layers of the system, including event logs produced by applications, the OS and IDSes, if present.
%As an instance, logs of the resource manager and of the scheduler can be used for collecting workload related information (time when an operation started, time when an operation ended, resources assigned for a certain task, etc.).
%Logs from the operating system can be collected for unearth possible malfunction of the system or attacks (e.g., login tentatives, data corruption).
%Logs generated by power distribution units contain information about power outages or spikes. 

% ***** END UNINA

\subsection{Adaptive Monitoring} 
%CIS: \\
%Analysis of the suitability of existing tools and technologies for adaptive monitoring:\\
%- Dedicated monitoring systems: can they be used adaptively?\\
%- General purpose data analysis systems\\
%- Hybrid Architectures

%-----INIZIO CIS ----

As shown in Figure~\ref{fig:framework}, the main idea is to adapt the monitoring by dynamically changing what raw data to collect and analyze, thus shaping at run time the resource utilization of the monitoring framework. 

All the monitoring systems described in the previous Section \ref{sec:collect} have been designed so as to allow users to plug-in custom modules, in order to extend their functionalities such to fit application-specific needs. The plug-in modules can be implemented so as to receive external commands that dynamically adapt their monitoring capabilities. 
%Further studies are needed to understand how much the openness of such a solution could fit the kind of adaptability we foresee.
%We are investigating the possibility of designing an adaptive monitor, that adapts the granularity of the monitoring activities in the different areas of the system depending on the alarms being raised \textbf{CIS-Avete Riferimenti?}. More precisely, \textbf{tutti: controllare che sia congruente con quanto detto al meeting)} 
An initial proposal to reduce the complexity (i.e. the quantity of collected data, hence required storage space and processing resources) is to define two different monitoring layers.
%, which we call coarse-grained layer and fine-grained layer. 
By default, the monitoring system operates in a \textbf{coarse-grained layer} collecting a limited number of variables, causing a high False Alarm Rate, but also a low Missed Detection Rate. 
In this configuration, the system acts as a very suspicious monitor which observes a reduced set of indicators and that easily raises alarms. 
When the coarse-grained layer detects an alarm in a specific area of the system, it triggers a \textbf{fine-grained layer} for monitoring that specific area through an enlarged set of indicators, a finer granularity of data, possibly reducing the False Alarm Rate. 

%The monitoring infrastructure should be designed taking into considerations, amongst the many:
%\begin{itemize}
%\item The considered attack model that characterizes the monitored system and that the anomaly detector aims to identify. For example, note that such two-layers approach is particularly fitting for revealing anomalies that last through time, while it may not suffice for anomalies that verify in "one-spot".
%\item The amount of indicators, nodes, and in general data monitored per unit of time, and the amount of resources required for the two monitoring layers.
%\item The expected occurrence of attacks and false alarms (both in time and in space i.g., in different areas of the critical infrastructure), that impacts the number of activations of the fine-grained monitoring and the distribution of such activations both in time and in space.
%\end{itemize}

The two-layers approach may lead to two different design solutions to be explored:
\begin{itemize}
\item The monitoring infrastructure is created with sufficient spare resources that are used to activate the fine-grained layer (we call this \emph{overprovisioning} approach). 
%This approach is feasible in case it is acceptable to have such amount of spare resource.
\item The monitored system is created such that it does not have spare resources, or with a very limited number of spare resource. The activation of the fine-grained layer in a certain area of the critical infrastructure requires to reduce the monitoring activity in some other areas (we call this the \emph{downgrade} approach).
\end{itemize}
The first approach leads to unused resources and the number of possible concurrent activations of the fine-grained layers 
%(in different areas of the critical infrastructure) 
is limited by the amount of spare resources. 
The second approach has no spare resources, but the downgrade of the monitoring activity risks to expose the system, leading to a not sufficient level of protection and/or to an unacceptable rate of False Alarm Rate.
% (which could cause a cascade of false alerts and consequent activations of fine-grained monitoring).
Also, the two approaches could be merged trying to take advantage of both of them. 

Clearly, the selection of the right approach and its tuning require to understand the distribution and temporal persistence of anomalies in the system. 
This is relevant to understand the expected frequency of fine-grained layer activations, and the extent to which it is possible to reduce the monitoring resolution without significantly affecting the detection of threats.

%Obviously, the selection of the approach and its tuning require to attentively consider the coverage offered by the anomaly detector and the possible burden on resources and existing tools are to be evaluated to understand how much the openness of such solutions could fit the kind of adaptability we foresee.
%Understanding which monitoring infrastructure is preferable for a target system depends on several parameters, for example the ones mentioned at the beginning of this section. 
%Deciding and evaluating the monitoring infrastructure requires the support of feedback from i) the execution of the case study, ii) the attack model and attackers profile, and iii) the definition of the anomaly detection algorithm, for example respectively i) to understand the resources required for each activation of the fine-grained layer, ii) to identify the Alarm Rate, and the topological distribution of Alarms in the different parts of the system, and iii) to estimate the False Alarm Rate (which still requires the activation of the fine-grained monitoring) for the expected workload. We are aware that there are known approaches to system, dependability and performability modeling, for example, \cite{meyer1980evaluating}, \cite{gaonkar2008scaling} that can result as useful guides for the selection of the best monitoring infrastructure for target system that we explore in Section \ref{sec:casestudy}.

On the other side of the spectrum, general purpose data analysis systems, which include a large family of tools like rule engines (e.g. Drools\footnote{\url{http://www.jboss.org/drools/}}), map reduce frameworks (e.g. Hadoop\footnote{\url{http://hadoop.apache.org/}}) and complex event processing systems (e.g. Esper\footnote{\url{http://esper.codehaus.org/}} and Storm\footnote{\url{http://storm.incubator.apache.org/}}) can be integrated with data gathering and diffusion platforms (e.g. multicast and publish/subscribe middleware) to create ad-hoc monitoring solutions. 
Nevertheless, the flexibility of this approach comes at a cost: most of these solutions must be designed and developed from scratch since often strictly tied to their initial target environment.

A possible solution we foresee for the Monitoring Adapter block is represented by an hybrid approach, where existing monitoring systems and general purpose data analysis tools are mixed and deployed in such a way to maximize their effectiveness in reaching the desired adaptability goals. Monitoring systems could locally analyze and observe specific subsystems to provide more high level information to data analysis tools for correlation with information provided by other different sources. The complexity involved in mixing these approaches together, however, remains to be studied.

%-------UNIFI ends-----%

%UNIFI: \\
%evaluating the ``cost'' of different approaches by means of (stochastic) models\\
%- how many resources do we need when fining the grain?\\
%- using of all resources vs. reserving part of resources\\
%- importance of scaling resources
%A scheduler to govern the data collection layer, i.e., activate/deactrivate data sources and increase/decrease monitoring frequency. Definition of possible approaches to vary the grain. Use of models to build categories of approaches.

\subsection{Data Analysis}
% ***** START UNINA

This component analyzes the data and provides as outputs information about
(i) how to adapt the grain of the monitoring,
%i) how to adapt the monitoring (e.g., by refining the granularity due to estimated upcoming events or by increasing the grain since threats appear overcome) and 
(ii) what protection actions should be performed on the CI.
Starting from our past experiences on attack modeling and data analysis, we consider the following functional blocks.

\textbf{Data Processing}. Collected raw data typically contain useless or redundant information that can undermine the goodness of performed analysis \cite{pecchia12}. The first analysis step to be performed is thus to polish raw data, adopting filtering or event coalescence techniques, such as the ones analyzed in \cite{dimartino12}.
% ***** END UNINA

%%%%%%%%%% INIZIO UNICAL %%%%%%%%%
\textbf{Attack Modeling}. This functional block provides tools
to define and statically analyze attack models. 
The attack model used in this block must be capable of:
%\begin{enumerate}
%\item 
(i) providing a high degree of flexibility in representing many different security scenarios in a compact way;
%\item 
(ii) allowing the specification of various kinds of constraints (e.g., temporal) on
possible attacks;
%\item 
(iii) representing attack scenarios at different abstraction levels, allowing to ``focus''
the conformance checking task in various ways.
%\end{enumerate}
Typed temporal graph-based attack models~\cite{GKR13} appear to be good options
for the above requirements. They are rich in terms of temporal
constraints that can be expressed. In addition, it is relatively easy
to handle the definition of generalization/specialization hierarchies among event types.
%Moreover, many interesting static analysis
%problems arise with such attack models. For instance, 
%it is important to be able to efficiently check the consistency of a given
%attack model when many temporal constraints are defined on it.
%A subset of the constraints can be contradictory~\cite{GKR13} and the system
%must be capable of informing its users about such situations.
%Another important static analysis problem is that of establishing
%the minimality of an attack model, in order to decrease the average size
%of the models whose conformance is being checked.

\begin{figure}[t!]
\centering
\includegraphics[scale=0.212]{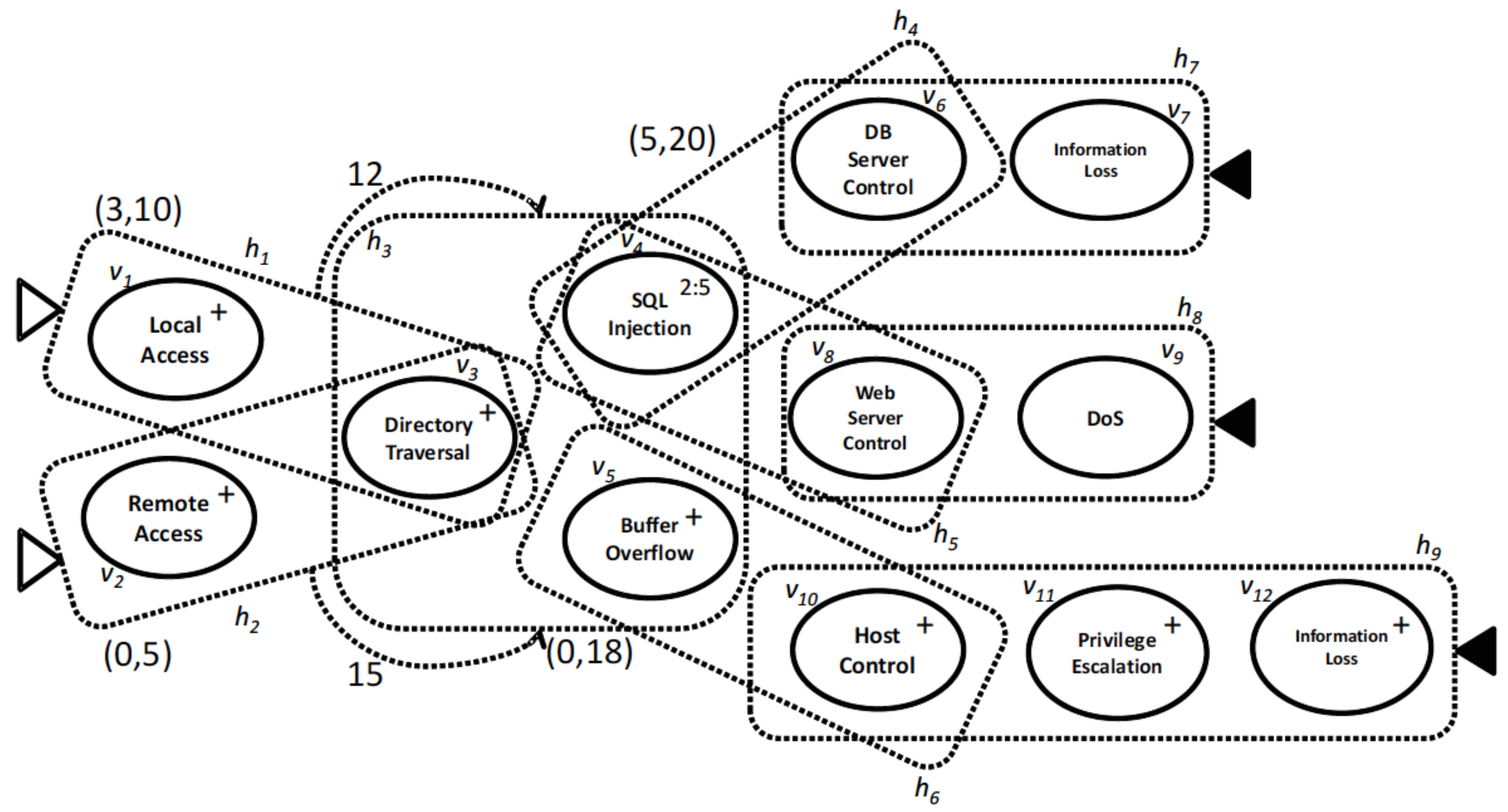}
\caption{\small{Example graph-based attack model}}
\label{fig:att-model}
\vspace{-0.5cm}
\end{figure}

By way of temporal graph-based model example, consider the hypergraph shown in
Fig.~\ref{fig:att-model}. Here it is assumed that the log is a sequence of tuples that represent 
high-level actions corresponding
to types of possible security exploits -- such logs can be built on-line from operational data. 
Actions are depicted with  plain circles ($v_i$), while (hyper-)edges are depicted with dotted circles ($h_i$).
As an instance, according to the semantics given in~\cite{GKR13},  $h_1$ is a start hyperedge (indicated with a white arrow) so an attack can begin with it.
The vertex labeled \emph{Local Access} requires the presence of a group of log tuples
with one ore more tuples of type \emph{Local Access} (cardinality constraint ``+''); 
the same applies to the \emph{Directory Traversal} vertex.
The hyperedge itself represents an association between the two vertices, with a temporal constraint of $(3,10)$ time points for the log segment. Hyperedge $h_3$ requires, in any order: (\emph{i}) one or more \emph{Directory Traversal} tuples; (\emph{ii}) between 2 and 5 \emph{SQL Injection} tuples; (\emph{iii}) one or more \emph{Buffer Overflow} tuples. The same applies to other hyperedges, such as $h_4$ and $h_7$. In particular, since $h_7$ is a terminal hyperedge (indicated with a black arrow), an attack can end with it.

 \textbf{Conformance Checking}. The main purpose of this functional block 
is that of detecting attack instances in sequences of logged events by checking the conformance 
of logged behavior with the given set of attack models.
The main requirement of this block is obviously scalability. In real-world critical
infrastructure protection scenarios, in fact, logged events are
streamed into the system on-line and, ideally, we would like to raise an alert as soon
as an event with a ``criticality'' above the threshold is logged.
It is therefore important to define appropriate data structures that ensure 
fast access to the relevant information, as well as suitable algorithms that are
tightly coupled with such structures in order to ensure the fast detection of
an attack~\cite{IRI07,TKDE13}.
Moreover, it is important to identify conditions that make the problem
tractable from a theoretical point of view. One possibility is that of imposing specific limitations to the structure
of the allowed models. In fact, recent work on the detection instances of temporal automaton-like
models in sequences of logged events~\cite{ESORICS11,CISIM11,TKDE13} has shown that acceptable
detection times in real-world cases can be obtained by limiting the number of partial solutions
through a form of early filtering based on temporal constraints. Finally,
the parallelization of both the data structures and the 
conformance checking algorithms (see, e.g.,~\cite{TKDE14}) appears mandatory when we target
big data for security protection.
%%%%%%%%%% FINE UNICAL %%%%%%%%%

\begin{figure}[t!]
\centering
\includegraphics[scale=0.87]{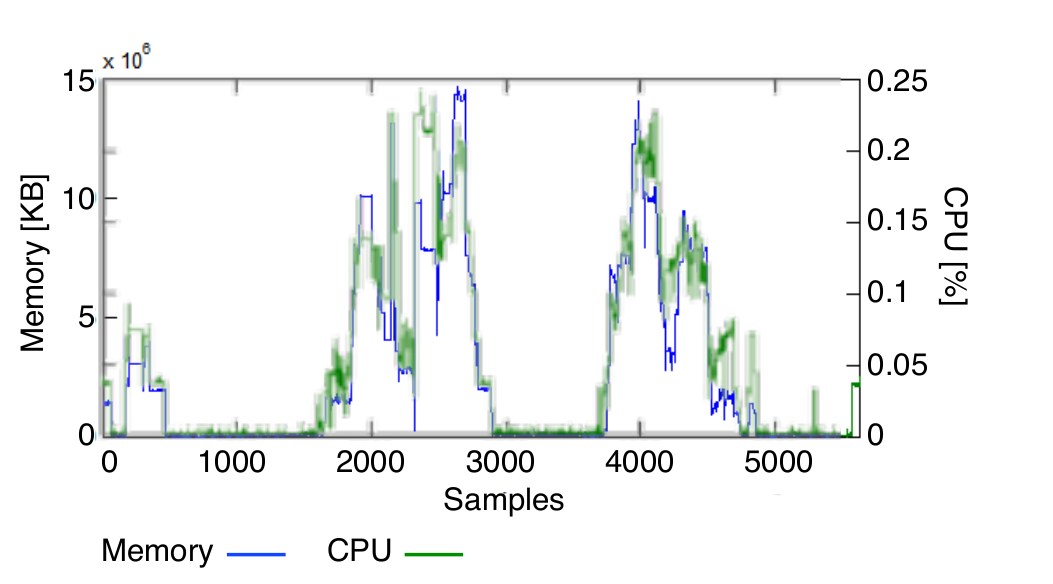}
\caption{\small{Invariant relationship between Memory and CPU percentage in a supercomputer.}}
\label{fig:invariants}
\vspace{-0.5cm}
\end{figure}

% ***** START UNINA
\textbf{Invariant-based Mining}. Invariants are properties of a system that are guaranteed to hold for all of its executions. If those properties are found to be violated (or broken) while monitoring the execution of the system, it is possible to trigger alarms useful to undertake immediate protection actions. As an example, figure \ref{fig:invariants} shows a relationship between the memory and CPU usage discovered from workload traces of the data center discussed in Section \ref{sec:casestudy}. Several studies have confirmed that is possible to discover invariants from real-world complex systems \cite{sharma_dsn13,lou_usenixatc10}. However, in our case the challenge is to discover invariant relationships in the big data collected from the CI. The \textit{Invariant-based Mining} block intends to face this issue,  performing two tasks: i) to automatically mine invariants from collected data streams using autoregressive models, and ii) to detect at runtime when invariant relationships are broken, to trigger immediate action. A preliminary application of the approach on real data collected from a production cloud software system has proven its feasibility and usefulness to discover execution deviations and SLA violations \cite{sarkar14}.

\textbf{Bayesian Inference}. Security monitors usually produce a large number of false alerts. A Bayesian network approach can be used on the top of the CI to correlate alerts coming from different sources and to filter out false notifications. This approach has been successfully used to detect credential stealing attacks \cite{pecchiaSRDS}. Raw alerts generated during the progression of an attack, such as user-profile violations and IDS notifications, are correlated trough a Bayesian network to pinpoint misuse performed by compromised users. The approach was able to remove around 80\% of false positives (i.e., not compromised user being declared compromised)
without missing any compromised user.

\textbf{Fuzzy Logic}.
Statistical methods cause a lot of false alarms.
This is due to the difficulty in defining exact and crisp rules describing when an event is an anomaly or not.
Boundaries between the normal and the anomaly behavior of a system are not clear and deciding the degree of intrusion at which an alarm is to be raised may vary in different situations \cite{feizollah_ccis13}.
Fuzzy logic is derived from fuzzy set theory to deal with approximate reasoning rather than precise data, and efficiently helps to smooth the abrupt separation of normality and abnormality \cite{frattini_lncs12}.
Anomalous events can be described by means of linguistic variables characterized by linguistic terms $an_i$.
The degree of truth of an expression is not crisp (i.e. of the type \textit{``$event_j$ is anomaly $a_i$''} or \textit{``$event_j$ is not anomaly $a_i$''}), and the use of fuzzy linguistic variables allows one to express vagueness in measurements.
In some applications, 99.95\% for attack detection accuracy has been reached \cite{feizollah_ccis13}.

% ***** END UNINA

\subsection{Case Studies}
\label{sec:casestudy}

In the framework of the Research Project of National Interest (PRIN) ``TENACE - Protecting National Critical Infrastructures from Cyber Threats'', we plan to experiment the framework on data extracted from two real-world systems.

The first is the data center of the Italian Ministry of Economic and Finance (MEF), which represents an important CI because it manages a wide range of software, spanning from very large applications with millions of end-users, such as those for the consumer credit support, up to small and very mission critical applications, such as those for managing the auctions of Italian Government Bonds and Treasury bills, and those for monitoring the government securities market (MTS).
In its architecture each rack is organized in up to five sub-racks. Each sub-rack can include up to sixteen blade servers and is connected to the datacenter network through four switches. A probe is connected to each switch in order to monitor flowing network traffic. Two smart PDUs are connected to each sub-rack to gather information about energy consumption. This configuration allows to enforce non-intrusive monitoring and to consider the system as a blackbox.

% ***** START UNINA

The second is the S.Co.P.E. supercomputer,  a scientific data center at the University of Naples Federico II and Italian Institute of Nuclear Physics (INFN). It is equipped with a monitoring system that collects data similar - for type and amount - to those collected by data centers of real CIs.
 %for data management and analysis, such as, the Advanced Simulation and Computing (ASC)\footnote{\url{http://nnsa.energy.gov/asc}} facilities of the U.S. National Nuclear Security Administration (NNSA).
%Examples are supercomputers used for military purposes or for space missions.
%High Performance Computing often supports Critical Infrastructures for data management and analysis. Examples are supercomputers used for military purposes or for space missions.
%In the U.S., the National Nuclear Security Administration (NNSA) relies on high performance computing for Advanced Simulation and Computing (ASC)\footnote{\small{http://nnsa.energy.gov/asc}}. Hence, it is representative and interesting to target a supercomputer to have a clear understanding on the type and amount of data that can be collected.
%In Italy, ... aggiungere esempio italiano
%As a supercomputer case study, we consider S.Co.P.E., a scientific data center at the University of Naples Federico II and Italian Institute of Nuclear Physics (INFN).
S.Co.P.E. mainly runs scientific batch jobs and also acts as a Tier-2 resource of the Worldwide LHC Computing Grid (WLCG)\footnote{The  Large Hadron Collider computing grid, http://wlcg.web.cern.ch}.
It is composed of 512 servers, each equipped with $2$ quad core CPUs and $32$ GB of memory.
For jobs queueing and scheduling and resource management, S.Co.P.E. uses Maui/TORQUE.
The monitoring system collects performance and environment data (adopting Ganglia extended with created additional scripts), and logs, producing about 0.7 GB of data per day.
Logs from Torque Resource Manager are used for collecting jobs' related data. Logs from the operating systems can be used for collecting anomaly related data.

% ***** END UNINA

\section{Closing Remarks and Open Issues}
%ALL: please add a description of other possible open issues you envisage.
% ***** START UNINA

Field data represent a rich source of information for improving the security monitoring and protection of future critical infrastructures. Existing monitoring technologies already offer the possibility of collecting different types of data, such as performance, environmental and operational data. The idea of collecting these types of heterogenous data, and analyze them trough a combination os state-of-art attack modeling and data analysis techniques, is promising to improve the accuracy of detection and drive the adaptation of the monitoring itself. In addition, the availability of existing analysis approaches and open, configurable monitoring tools represents a good start for the viability of the proposed framework. 

However, the achievement of envisioned research objectives requires to face many open issues, such as:
\begin{itemize}
\item lack of publicly available data sets and ground truth. Such data sets are vital for the validation of approaches like the one proposed in this paper. Some datasets are outdated (such as DARPA\footnote{http://www.ll.mit.edu/mission/communications/cyber/CSTcorpora/ideval/data/}) or unlabeled (such as, iCTF\footnote{http://ictf.cs.ucsb.edu}), while others target standard IT systems (such as UNB iSCX\footnote{http://www.iscx.ca/datasets}). To date, there are no datasets available for CIs.
\item difficulty of performing long-running tests on real-world systems (or representative reproductions). These tests are very useful to improve the understanding of phenomena and to produce realistic datasets. However, honeypot-like approaches cannot be adopted in the case of CIs, due to the possibility of physical damages as a consequence of an attack. Innovative controlled environments are to be created, involving the expertise and equipment of CIs stakeholders.
\item need of strategies for changing the monitoring configuration at runtime on the basis of some predefined logic, without being forced to stop and restart the services in charge of gathering and analyzing data. This is important because it provides large freedom in adapting the monitoring without interrupting related services.
\item urgency of scalable solutions to combine the outputs generated by the different data analysis techniques, as the ones envisaged in the framework. Further research is in order, involving the different views and know-how of the researchers active in these fields.
% ***** END UNINA
\end{itemize}	
Hence, the path towards industry-ready solutions calls for further joint industry-academia efforts, involving major players and stakeholders, to take real advantage of big data for the security monitoring of future critical infrastructures.

%Commento: dividerei in \emph{Technical challenges} come ad esempio le "scalable solutions" di sopra, e quelle aggiunte sotto, e qualcosa tipo \emph{Operative challenges} per indicare le difficolta' non specificatamente "tecniche" (mancanza di dati, tempo per la raccolta, difficoltà di avere accesso a honeypots, ...).

%From a technical point of view, it is also needed to:
%\begin{itemize}
%\item Define the attack model for the anomaly detection monitor, and understand how anomalies are generated wrt the selected attacks. This shall be fundamental for defining the policies of the two-layers monitoring approach, including the required grained for the monitoring system. Also fundamental is to understand the temporal persistence of an anomaly, to define the required deadlines for the timely activation of the fine-grained monitoring layer.
%\item Understand the frequency and distribution of anomalies in the system, and also of false alarms, to understand the appropriate configuration of the monitoring system. This will be relevant to understand what is the expected frequency of activations of the fine-grained layer, and the extent to which it is possible to reduce the monitoring resolution without significantly affecting the coverage of the anomaly detector.
%\end{itemize}


\section*{Acknowledgment}
This work has been supported by the TENACE PRIN Project (n. 20103P34XC) funded by the Italian Ministry of Education, University and Research.

\bibliographystyle{IEEEtran}
\bibliography{IEEEabrv,references}

\end{document}